\documentclass[a4paper,11pt]{article}
\pdfoutput=1 

\usepackage{jheppub} 

\usepackage[T1]{fontenc} 

\usepackage[multiple]{footmisc}

\usepackage{graphicx}
\usepackage{amssymb, amsmath,mathtools,amsthm}
\usepackage{hyperref}
\hypersetup{
    colorlinks=true,
    linkcolor=black,          
    citecolor=black,        
    filecolor=black,      
    urlcolor=black          
}

\def\calv{\mathcal{V}}

\newcommand{\beq}{\begin{equation}}
\newcommand{\eeq}{\end{equation}}

\newcommand{\lie}{\mathcal{L}}
\newcommand{\diff}{{\rm Diff}}
\newcommand{\sdiff}{{\rm Diff_{\rm vol}}}
\newcommand{\vect}{{\rm Vect}}
\newcommand{\svect}{{\rm SVect}}
\newcommand{\lieg}{\mathfrak{g}}

\newcommand{\be}{\mathbf{e}}

\title{\boldmath Near-horizon BMS symmetries as fluid symmetries}

\author[a]{Robert F. Penna}
\affiliation[a]{Center for Theoretical Physics, Columbia University,\\
New York, New York 10027, USA}
\emailAdd{rp2835@columbia.edu}

\abstract{

The Bondi-van der Burg-Metzner-Sachs (BMS) group is the asymptotic symmetry group of asymptotically flat gravity.  
Recently, Donnay et al. have derived an analogous symmetry group acting on black hole event horizons. 
For a certain choice of boundary conditions, it is a semidirect product of $\diff(S^2)$, the smooth diffeomorphisms of the two-sphere, acting on $C^\infty(S^2)$, the smooth functions on the two-sphere.  We observe that the same group appears in fluid dynamics as symmetries of the compressible Euler equations.  We relate these two realizations of $\diff(S^2)\ltimes C^\infty(S^2)$ using the black hole membrane paradigm.  We show that the Lie-Poisson brackets of membrane paradigm fluid charges reproduce the near-horizon BMS algebra.  The perspective presented here may be useful for understanding the BMS algebra at null infinity.

}

\begin{document} 
\maketitle
\flushbottom

\section{Introduction}
\label{sec:intro}

The Bondi-van der Burg-Metzner-Sachs (BMS) group is the asymptotic symmetry group of asymptotically flat gravity \cite{1962RSPSA.269...21B,1962RSPSA.270..103S}.  It is similar to the Poincar\'{e} group, except the translation subgroup is enlarged to an infinite dimensional group of angle-dependent ``supertranslations.''  For certain choices of boundary conditions, the Lorentz subgroup is enlarged as well, to an infinite dimensional group of superrotations.  Depending on the choice of boundary conditions, the superrotations comprise either the infinitesimal conformal transformations of the two-sphere \cite{2010PhRvL.105k1103B,Barnich:2011ct} or else the smooth diffeomorphisms of the two-sphere, $\diff(S^2)$ \cite{2014PhRvD..90l4028C,2015JHEP...04..076C}. 

The BMS group describes the physical symmetries of asymptotically flat gravity.  If asymptotically flat gravity is holographic, then the BMS group will govern its dual (just as the conformal group governs dual descriptions of asymptotically anti-de Sitter gravity).  So it is of interest to look for (non-gravitational) physical systems governed by BMS symmetry.  The case for which the superrotations comprise the infinitesimal conformal transformations of the two-sphere was recently explored by \cite{Kapec:2016jld}.
It is also of interest to consider the case for which the superrotations comprise $\diff(S^2)$.    In this case, the BMS group is closely related to the physical symmetries of fluid dynamics.  In particular, Marsden et al. \cite{marsden1984semidirect,marsden1984reduction} showed long ago that the semidirect product $\diff(M)\ltimes C^{\infty}(M)$ governs the compressible Euler equations on $M$.  The goal of the present paper is to relate this observation to BMS symmetry.  The idea that gravity and fluids are linked goes back to the black hole membrane paradigm  \cite{Damour:1979wya,1982mgm..conf..587D,Price:1986yy,1986bhmp.book.....T,1998PhRvD..58f4011P,2015PhRvD..91h4044P} and the fluid/gravity correspondence \cite{Hubeny:2011hd}.   We derived BMS conservation laws from membrane paradigm dynamics in \cite{Penna:2015gza} and the relationship between BMS and fluids was also pursued by \cite{Eling:2016xlx,Eling:2016qvx} from a somewhat different perspective.  

The BMS group at null infinity is not quite the same as the symmetry group of an ordinary compressible fluid on $S^2$.  Both can be realized as semidirect products of $\diff(S^2)$ acting on $C^\infty(S^2)$, but the actions appearing in the semidirect products are different.  BMS  supertranslations at null infinity transform as negative half-densities \cite{Barnich:2011ct,2015JHEP...04..076C}, whereas the ``supertranslations'' of fluid dynamics transform as ordinary functions \cite{marsden1984semidirect}.

However, the semidirect products become identical if we consider the near-horizon BMS group.  The near-horizon BMS group is defined at black hole event horizons as the group of near-horizon Killing vectors preserving a set of near-horizon boundary conditions\footnote{Just as at null infinity, the near-horizon superrotations comprise either the infinitesimal conformal transformations of $S^2$ or $\diff(S^2)$, depending on the choice of boundary conditions.  We identify the near-horizon superrotations with $\diff(S^2)$.} \cite{Donnay:2015abr}.  Our main result is to recover the near-horizon BMS algebra by taking Lie-Poisson brackets of fluid charges.  We use the black hole membrane paradigm as a dictionary between the gravity and fluid sides of the story.   

Let us say a bit more about the infinite dimensional symmetries of fluids, beginning with the incompressible case for simplicity.  The incompressible Euler equations depend on the velocities but not the spatial labels of fluid elements.  This infinite dimensional ``particle relabeling'' symmetry gives an infinite number of conserved charges \cite{arnold1966geometrie,arnold1969hamiltonian,marsden1983coadjoint,marsden2002introduction}.  The configuration space of the incompressible Euler equations for fluid flow on a space\footnote{We assume $M$ is closed but the extension to manifolds with boundary is straightforward.} $M$ is $G=\sdiff(M)$, the set of volume-preserving diffeomorphisms of $M$.  A configuration, $\varphi\in G$, is a map from Lagrangian coordinates to Eulerian coordinates.  It is a map from the initial positions of the fluid parcels to their current positions.  The fluid's phase space is the cotangent bundle $T^*G$.  It is the space of pairs $(\varphi, V)$, where $V$ is the fluid's Lagrangian momentum.  Now $G$ acts on $T^*G$ on the left and right, and we have corresponding moment maps $T^*G\rightarrow \lieg^*$, where $\lieg^*$ is the dual of the Lie algebra of $G$.  The left moment map is the fluid's Eulerian momentum and the right moment map is the fluid's convected momentum.  The particle-relabeling symmetry of the incompressible Euler equations is reflected in the right-invariance of the fluid's Hamiltonian on $T^*G$.  As such, we have conservation of the right moment map and  the dynamics reduces to a Hamiltonian flow on $\lieg^*$.  The Poisson bracket on $T^*G$ descends to the Lie-Poisson bracket on $\lieg^*$.  We review this story in more detail in the Appendix.

The extension to compressible fluids is somewhat subtle, but one essentially replaces $G=\sdiff(M)$ with $\diff(M)\ltimes C^\infty(M)$ \cite{marsden1984semidirect,marsden1984reduction}.  The components (with respect to a basis of $\lieg$) of the right moment map now correspond to the fluid's convected momentum density and convected energy density.  We will use the black hole membrane paradigm to relate the conserved charges of the membrane paradigm fluid to the near-horizon BMS charges of \cite{Donnay:2015abr}.  We will show that taking Lie-Poisson brackets of the fluid charges reproduces the near-horizon BMS algebra. 

It would be interesting to understand if the BMS algebra at null infinity also has a realization as symmetries of a (perhaps somewhat unusual) fluid.  As noted earlier, the action of the superrotations on the supertranslations at null infinity does not seem to have appeared in ordinary fluid dynamics.  On the other hand, a version of the membrane paradigm can be defined at null infinity \cite{2015PhRvD..91h4044P}. The problem simplifies in three spacetime dimensions.  In three spacetime dimensions, the BMS group has the form $\diff(S^1)\ltimes \mathfrak{vir}$, where $\mathfrak{vir}$ is the Virasoro algebra with central charge $c=3/G$.  The charge algebra arises from a Lie-Poisson bracket \cite{2014JHEP...06..129B,Barnich:2015uva} and the boundary dynamics has a simple description \cite{Barnich:2012rz,Barnich:2013yka,Carlip:2016lnw}.   We will study the fluid interpretation of 3d flat gravity in a companion paper \cite{Penna:2017vms}.  In the future, it would be interesting to use the fluid perspective to understand the symmetry groups appearing at Rindler horizons \cite{Afshar:2015wjm}, with negative cosmological constant \cite{Afshar:2016wfy,Afshar:2016kjj}, and in more general theories of gravity (e.g., \cite{Setare:2016qob,Setare:2017mry}).

The remainder of this paper is organized as follows.  Section \ref{sec:star} describes a kind of near-horizon ``memory effect'' as a way to develop intuition for the near-horizon BMS group.  Section \ref{sec:horizon} describes our main result, the realization of near-horizon BMS charges as fluid Noether charges.  Section \ref{sec:discuss} discusses the relationship between the physical symmetries of fluids and the gauge symmetries of canonical general relativity more generally.  We note that asymptotic symmetries may generally be expected to form groupoids rather than groups.   The Appendix gives a mostly self-contained review of the infinite dimensional symmetries and Noether charges of the incompressible Euler equations.

\section{Plunging star}
\label{sec:star}

In this section, we consider the plunge of a small star into a Schwarzschild black hole.  The black hole remains Schwarzschild after the plunge.  However, the passage of the star causes a permanent shift in the spatial metric on the horizon and the synchronization of clocks at the horizon.  The initial and final black holes are related by near-horizon BMS transformations.  This gives a near-horizon version of the gravitational memory effect.  We discuss it here as a way to develop intuition for the near-horizon BMS group.  The plunging star problem was solved long ago by Suen, Price, and Redmount (SRP) \cite{Suen:1988kq}.  Our novelty is to reinterpret it in terms of near-horizon BMS transformations and memory.

Consider the radial plunge of a small star into the north pole of a Schwarzschild black hole.  The star-to-black hole mass ratio is a small parameter, $\epsilon=m/M$, and we restrict attention to polar coordinates $\theta\ll 1$ (we zoom in near the north pole).  
SRP worked out the tidal field of the plunging star, a gravitational analogue of the Li\'{e}nard-Wiechert potential.   From the tidal field they determined the shear, $\sigma_{AB}$, of the horizon. The result is
\beq\label{eq:shear}
\sigma_{\hat{\theta}\hat{\theta}} = -\sigma_{\hat{\phi}\hat{\phi}} = 
\begin{cases} 
      -\frac{\epsilon}{M\theta^2}e^{\kappa t} & t < 0 \\
      0 & t > 0,
   \end{cases}
\eeq
where $\kappa$ is the horizon's surface gravity.  The components of the shear are measured in a local frame, $(e^{\hat{t}},e^{\hat{r}},e^{\hat{\theta}},e^{\hat{\phi}})$, which is essentially the zero-angular momentum observer (ZAMO) frame.  The star hits the horizon at $t=0$ moving near the local speed of light.  Note that the shear starts growing before the star hits because of the teleological nature of the horizon.

The shear determines everything else.  It determines the expansion of the horizon via the Raychaudhuri equation, the extrinsic curvature via the Hajicek equation, and the spatial metric via the metric evolution equation.

\subsection{Metric}

\begin{figure}
\centering
\includegraphics[width=0.75\columnwidth]{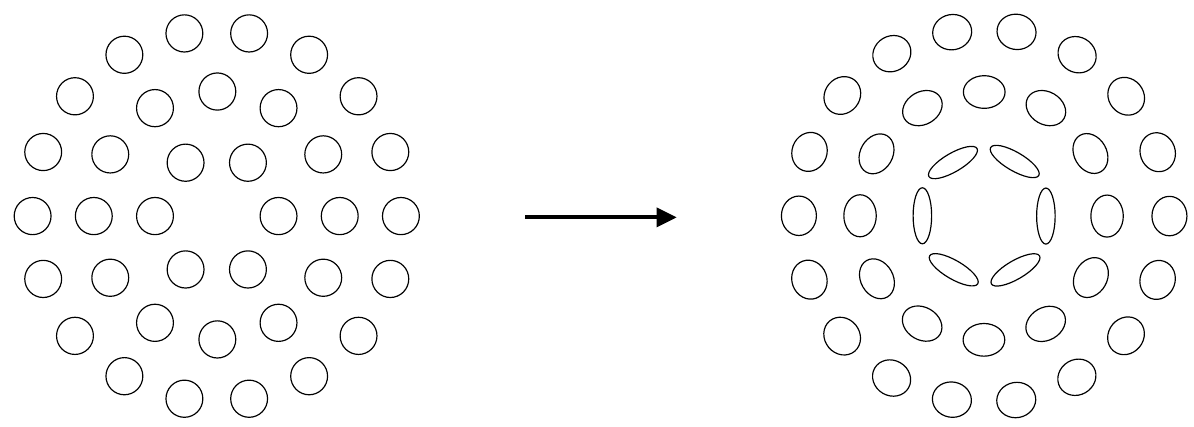}
\caption{We have zoomed in on the north pole.  Initially circular rings are squeezed along $\theta$ and stretched along $\phi$ by the passage of the star.}
\label{fig:memory}
\end{figure}

Long before the plunge, the spatial metric on the horizon is simply the round metric,
\beq\label{eq:dsi}
ds^2 = r_+^2 d\theta^2 + r_+^2 \sin^2\theta d\phi^2.
\eeq
We work in ``comoving coordinates,'' meaning $(\theta,\phi)$ are fixed to the null generators of the horizon.  Long after the plunge, the black hole is again Schwarzschild but the null generators do not return to their starting positions.  So the final spatial metric on the horizon is not the round metric, but instead
\beq\label{eq:dsf}
ds^2 = r_+^2\left(1-\frac{8\epsilon}{\theta^2}\right) d\theta^2 + r_+^2 \left(1+\frac{8\epsilon}{\theta^2}\right) \sin^2\theta d\phi^2.
\eeq
The initial and final metrics are related by the spatial diffeomorphism (``superrotation'')
\beq\label{eq:sr}
\theta \rightarrow \theta + \frac{4\epsilon}{\theta}.
\eeq
Figure \ref{fig:memory} gives a visualization: initially circular rings of test particles hovering near the horizon are distorted by the passage of the star.  It is a near-horizon analogue of the gravitational memory effect.

\subsection{Momentum density}
\label{sec:momentum}

The membrane paradigm assigns the horizon a stress-energy tensor, $t_{ab}=(Kh_{ab}-K_{ab})/(8\pi)$, where $K_{ab}$ is the extrinsic curvature and $h_{ab}$ is the $2+1$ dimensional horizon metric \cite{1986bhmp.book.....T}.  
The momentum density of the horizon is $\Pi_A\equiv t^{\hat{t}}_A$. 

The initial momentum density of the black hole is
\beq
\Pi_A = 0.
\eeq
As the star plunges, the horizon's momentum density evolves according to the Damour-Navier-Stokes equation (essentially the projection of the $\hat{r}A$-component of the Einstein equation onto the horizon).  Long after the plunge, the momentum density is
\begin{align}
\Pi_{\hat{\theta}} &= \frac{1}{8\pi}\frac{\epsilon}{\kappa M^2 \theta^3},\\
\Pi_{\hat{\phi}} &= 0.
\end{align}
The initial and final momentum densities are related by the slicing transformation (``supertranslation'')\footnote{Near the north pole, the slicing transformation, $t\rightarrow t + f(\theta)$, looks like a low-velocity Lorentz transformation, $t \rightarrow t - \delta v_{\theta} \theta$, with velocity $\delta v_{\theta}=-f_{,\theta}$.  The change in the momentum density is $\delta \Pi_\theta = p (-\delta v_\theta) = pf_{,\theta}$, where $p=\kappa/(8\pi)$ is the membrane's pressure \cite{1986bhmp.book.....T}.}
\beq
t\rightarrow t + \frac{\epsilon}{M\kappa^2\theta^2}.
\eeq
So the passage of the star has caused a supertranslation and a superrotation of the horizon.

This gives a physical picture for the action of supertranslations and superrotations at the horizon for the special case of a small star plunging into a Schwarzschild black hole.  In the next section, we consider perturbed black holes more generally, and derive the near-horizon BMS charge algebra using the membrane paradigm.

\section{Near-horizon BMS symmetries as fluid symmetries}
\label{sec:horizon}

Following Donnay et al. \cite{Donnay:2016ejv}, we write the near-horizon metric in Gaussian null coordinates \cite{Moncrief:1983xua,chrusciel},
\beq
ds^2=-2\kappa\rho dv^2 + 2d\rho dv + 2\theta_A \rho dv dx^A +\Omega\gamma dzd\bar{z} + \dots,\label{eq:metric}
\eeq
where $v$ is Eddington-Finkelstein advanced time, the $x^A=(z,\bar{z})$ are fixed to null generators of the horizon, and $\gamma=4/(1+z\bar{z})^2$.  The metric functions, $\theta_A=\theta_A(v,z,\bar{z})$ and $\Omega=\Omega(v,z,\bar{z})$, are independent of the radial coordinate, $\rho$.  For now we assume $\kappa=\kappa(v,z,\bar{z})$, but we will soon set $\kappa={\rm constant}$.  The horizon is at $\rho=0$.

To make contact with the membrane paradigm, we recast the metric in a timelike frame.  We introduce a timelike tetrad,
\begin{align}
\be^\rho		&= d\rho, \\
\be^t 		&= dv - \be^{r^*}, \\
\be^{z'} 		&= dz + \be^{r^{**}}, \\
\be^{\bar{z}'}	&= d\bar{z} + \be^{r^{***}},
\end{align}
with $e^{r^*}$, $e^{r^{**}}$, and $e^{r^{***}}$ chosen so as to eliminate $g_{t\rho}$, $g_{\rho z'}$ and $g_{\rho \bar{z}'}$ from the metric.  This requires
\begin{align}
\be^{r^*}	&= \frac{\gamma \Omega}{2\kappa\rho}(\gamma\Omega+2\rho \theta_z \theta_{\bar{z}}/\kappa)^{-1} \be^\rho,\\ 
\be^{r^{**}} &= \frac{\theta_{\bar{z}}}{2\kappa\rho}(\gamma\Omega+2\rho \theta_z \theta_{\bar{z}}/\kappa)^{-1} \be^\rho,\\ 
\be^{r^{***}} &= \frac{\theta_z}{2\kappa\rho}(\gamma\Omega+2\rho \theta_z \theta_{\bar{z}}/\kappa)^{-1} \be^\rho.
\end{align}
The metric in the timelike frame is 
\beq
ds^2 = 	- \alpha^2 \thinspace \be^{t} \otimes \be^{t}
		+\be^\rho \otimes \be^{r^*}
		+2\rho \theta_{A'} \be^{t} \otimes \be^{A'}
		+\Omega\gamma \be^{z'} \otimes \be^{\bar{z}'},
\eeq
where $\alpha=\sqrt{2\kappa\rho}$ is the lapse.
Further define
\begin{align}
\be^{\hat{t}} &= \alpha \be^t, \\
\be^{\hat{\rho}} &= \sqrt{g_{\rho\rho}} \be^\rho, \\
\be^{\hat{x}} &= \frac{1}{2}\sqrt{\gamma\Omega}(\be^{z'}+\be^{\bar{z}'})
		+\frac{1}{2}\frac{\rho (\theta_{z}+\theta_{\bar{z}})}{\sqrt{\gamma\Omega}}\be^{t}, \\
\be^{\hat{y}} &= \frac{1}{2i}\sqrt{\gamma\Omega}(\be^{z'}-\be^{\bar{z}'})
		+\frac{1}{2i}\frac{\rho(\theta_{\bar{z}}-\theta_{z})}{\sqrt{\gamma\Omega}} \be^{t}.
\end{align}
This is an orthonormal timelike frame.  The metric in this frame is simply
\beq
ds^2 = - \be^{\hat{t}}\otimes \be^{\hat{t}} + \be^{\hat{\rho}}\otimes \be^{\hat{\rho}}
	+\be^{\hat{x}}\otimes \be^{\hat{x}} + \be^{\hat{y}} \otimes \be^{\hat{y}}.
\eeq

Fix a surface, the ``stretched horizon,'' at small but nonzero $\rho$, with unit outward normal $n=e^{\hat{r}}$.  The induced metric on the stretched horizon is
\beq
h_{\mu\nu} = g_{\mu\nu} - n_\mu n_\nu.
\eeq
The membrane paradigm assigns the stretched horizon a surface stress-energy tensor,
\beq
t_{\mu\nu} = \frac{1}{8\pi G}(K h_{\mu\nu}-K_{\mu\nu}),
\eeq
where ${K^\mu}_\nu = h^\delta_\nu \nabla_\delta n^\mu$ is the extrinsic curvature of the stretched horizon and $K={K^\mu}_\mu$.  The stress-energy tensor is dictated by the Israel junction condition.  It is the stress-energy tensor required to terminate the gravitational field at the stretched horizon.  The Einstein equations enforce the conservation laws
\beq
{{t_\mu}^\gamma}_{|\gamma} = h_{\mu \delta} h^\nu_\gamma \nabla_\nu t^{\delta\gamma} = 0.\label{eq:consv}
\eeq

Now for a weakly perturbed black hole, Price and Thorne \cite{Price:1986yy} have shown that it is always possible to arrange $\kappa={\rm constant}$ (by applying a reparametrization of $v$).  From now on, we will assume this has been done as it simplifies the analysis considerably.  In this case, the fluid's energy density is
\beq
\Sigma_H \equiv \alpha t^{\hat{t}\hat{t}} = -\frac{\partial_v \Omega}{8\pi G \Omega} = \frac{\theta_H}{8\pi G},
\eeq
where $\theta_H=\alpha \gamma^\mu_\nu \nabla_\mu n^\nu = \partial_\nu \Omega/\Omega$ is the expansion scalar.  The evolution of $\Sigma_H$ is governed by the $\mu=t$ component of \eqref{eq:consv}, which gives
\beq
\partial_v \Sigma_H + \Sigma_H \theta_H  + p_H \theta_H -\zeta_H \theta_H^2 = 0,
\eeq
where $p_H =\alpha^{-1}{K^\mu}_\nu U_\mu U^\nu/(8\pi G) = \kappa/(8\pi G)$ is the fluid's pressure and $\zeta_H = -1/(16\pi G)$ is its bulk viscosity.

The fluid's momentum density is 
\beq
\Pi_A = t^{\hat{t}}_A = \frac{\theta_A}{16\pi G}.  
\eeq
Its evolution is governed by the $\mu=z,\bar{z}$ components of \eqref{eq:consv},
\beq
\partial_v \Pi_A + \Pi_A \theta_H - \zeta_H \nabla_A \theta_H = 0,
\eeq
which is sometimes called the Damour-Navier-Stokes equation.  

\subsection{Non-expanding black hole}
 
For a non-expanding black hole ($\theta_H = 0$), the Damour-Navier-Stokes equation is simply $\partial_v \Pi_A = 0$ and $\Pi_A$ is exactly conserved.  In this case, there is a doubly-infinite family of conserved scalar charges,
\beq
Q_{X,f} = \int dzd\bar{z}\sqrt{\gamma}\Omega(fp_H-X^A\Pi_A) 
	= \frac{1}{16\pi G}\int dzd\bar{z}\sqrt{\gamma}\Omega(2f\kappa-X^A\theta_A).
\eeq
There is a charge for each $X\in \vect(S^2)$ and $f\in C^\infty(S^2)$.   As in ordinary fluid dynamics, we view the pair $(X,f)$ as an element of $\lieg$, the Lie algebra of $G=\diff(S^2)\ltimes C^\infty(S^2)$, and we view the charges themselves  as elements of the dual, $\lieg^*$.  The space $\lieg^*$ is a Poisson manifold with respect to the Lie-Poisson bracket.  The Lie-Poisson bracket acts on functions $F,G:\lieg^*\rightarrow \mathbb{R}$ according to
\beq
\{F,G\}(m) = \left\langle m, \left[\frac{\delta F}{\delta m},\frac{\delta G}{\delta m}\right] \right\rangle.
\eeq
So the algebra of fluid charges is
\begin{align}
\{Q_X,Q_Y\} &= Q_{[X,Y]},\label{eq:q1}\\
\{Q_X,Q_f\} &= Q_{X\cdot f},\label{eq:q2}\\
\{Q_f,Q_g\} &= 0.\label{eq:q3}
\end{align}
This matches the near-horizon BMS charge algebra of Donnay et al. \cite{Donnay:2016ejv} (for a certain choice of boundary conditions).  The relationship with fluid dynamics provides a physical interpretation for the charges, which may be useful for understanding asymptotic symmetries in more general situations.

\section{Discussion}
\label{sec:discuss}

BMS symmetries are the physical symmetries of asymptotically flat gravity.  They are closely related to the gauge symmetries of general relativity.  As we explain in this section, once one knows the gauge symmetries of general relativity and the physical symmetries of compressible fluids, it is not so surprising that the two are closely related.

In the canonical formulation of general relativity, spacetime is foliated into three-dimensional spatial slices.  The configuration on a slice is given by the spatial metric on the slice, $\gamma$, and its conjugate momentum, $\pi$.  Configurations cannot be freely specified.  They are subject to the energy and momentum constraints
\begin{align}
\mathbf{C}_{\rm en} &= -R(\gamma) + {\rm tr}_\gamma(\pi^2)-\frac{1}{2}({\rm tr}_\gamma\pi)^2 = 0,\label{eq:con1}\\
\mathbf{C}_{\rm mo} &= -2{\rm \thinspace div}_\gamma \pi = 0,\label{eq:con2}
\end{align}
where $R(\gamma)$ is the scalar curvature of $\gamma$.   Projecting against functions, $\varphi$, and vector fields, $X$, gives scalar-valued constraint functions
\begin{align}
C_\varphi &= \int_\Sigma \varphi \mathbf{C}_{\rm en} {\rm vol}_\gamma,\\
C_X &= \int_\Sigma X \cdot \mathbf{C}_{\rm mo} {\rm vol}_\gamma.
\end{align}
The Poisson brackets of the constraint functions are \cite{katz1962relativite,dewitt1967quantum,Blohmann:2010jd}
\begin{align}
\{C_X,C_Y\} &= C_{[X,Y]},\label{eq:c1}\\
\{C_X,C_\varphi\} &= C_{X\cdot \varphi},\label{eq:c2}\\
\{C_\varphi,C_\psi\} &= C_{\varphi\nabla_\gamma \psi - \psi\nabla_\gamma \varphi}.\label{eq:c3}
\end{align}
The appearance of $\gamma$ on the rhs of the third line means the structure constants are state dependent.  So the gauge symmetries of canonical general relativity do not form a Lie group.   They form a Lie groupoid\footnote{A groupoid is a category with inverses.  A group is a one-object groupoid.  So groupoids represent a more flexible notion of symmetry than groups \cite{mackenzie1987lie}.}, a point emphasized by \cite{Blohmann:2010jd}.

\begin{figure}
\centering
\includegraphics[width=0.75\columnwidth]{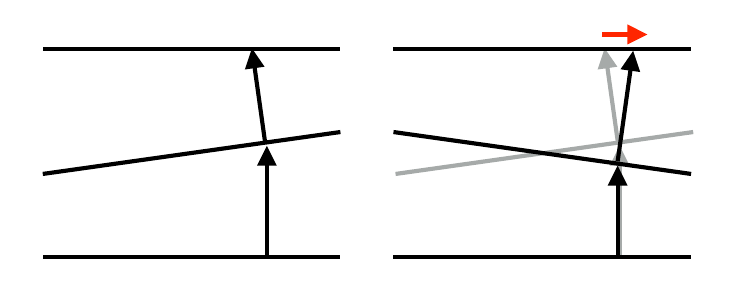}
\caption{Slicing transformations, $t\rightarrow t+f(x^A)$, do not commute.  After a slicing transformation, the normal of the new spatial slice is tilted.  The commutator of two infinitesimal slicing transformations is a spatial diffeomorphism.}
\label{fig:shift}
\end{figure}

Evidently, there is a close relationship between the gauge symmetries of canonical general relativity and the physical symmetries of compressible fluids in one less dimension.  The only difference between the fluid algebra \eqref{eq:q1}-\eqref{eq:q3} and the gauge algebroid \eqref{eq:c1}-\eqref{eq:c3} is in the third line.  This can be traced back to the fact that time-reparametrizations commute at black hole horizons (where the lapse is going to zero), but they do not commute in general.   The reason they do not commute in general is illustrated in Figure \ref{fig:shift}.  After a nonconstant slicing transformation, $t\rightarrow t+f(x^A)$, the normal of the new spatial slice is tilted.  So a sequence of slicing transformations depends on the order they are carried out.  The commutator of two infinitesimal slicing transformations is a spatial diffeomorphism.  This accounts for the rhs of \eqref{eq:c3}.

In asymptotically flat spacetimes, time-reparametrizations commute at null infinity.  This is because the gradients, $\nabla_\gamma \sim r^{-2}$, vanish at null infinity.  So the state-dependent terms on the rhs of \eqref{eq:c3} are not visible in the BMS group at either null infinity or the horizon.

However, the rhs of \eqref{eq:c3} does have an interesting role to play in asymptotically de Sitter spacetimes.  Now the lapse is growing at null infinity, and its growth cancels the decay in $\nabla_\gamma$.
In static coordinates, the metric of four dimensional de Sitter space is
\beq
ds^2 = -\left(1-\frac{r^2}{\ell^2}\right)dt^2 + \frac{dr^2}{1-r^2/\ell^2}+r^2d\Omega^2\label{eq:ds}.
\eeq
On surfaces of large but finite $r$, the de Sitter metric becomes 
\beq
ds^2 = \frac{r^2}{\ell^2} dt^2 +r^2d\Omega^2.
\eeq
This is the metric on ``stretched infinity.''

Now consider $\diff(S^2)\times C^\infty(S^2)$, where the first factor corresponds to spatial diffeomorphisms of $S^2$ and the second factor corresponds to slicing transformations, $t\rightarrow t+f$.  The commutator of two slicing transformations is 
\beq
[f,g] = N^2 (g\nabla_h f-f\nabla_h g),
\eeq
where $N=r/\ell$ is the lapse and $h$ is the metric on $S^2$.   At large $r$, the gradient is $\nabla_h= r^{-2} \nabla_{\hat{h}}$, where $\nabla_{\hat{h}}$ is the gradient of the rescaled metric $\hat{h}=h/r^2$.  So the commutator is
\beq
[f,g] = \frac{1}{\ell^2} (g\nabla_{\hat{h}} f-f\nabla_{\hat{h}} g).\label{eq:fg}
\eeq
For finite $\ell$, slicing transformations do not commute. 

\acknowledgments

This research was supported by a Prize Postdoctoral Fellowship in the Natural Sciences at Columbia University.  We are grateful to Geoffrey Comp\`{e}re,  St\'{e}phane Detournay, Lam Hui, Austin Joyce, Alberto Nicolis, Blagoje Oblak, and Rachel Rosen for comments.


\appendix

\section{Infinite dimensional symmetries of the Euler equations}
\label{sec:euler}

This appendix gives a mostly self-contained review of the infinite dimensional symmetries and Noether charges of the incompressible Euler equations, as background and motivation for the results in the main body of this paper.  See \cite{marsden1984semidirect} for the extension to compressible fluids.

The incompressible Euler equations are
\begin{align}
\partial_t v + (v\cdot \nabla)v = -\nabla p,\label{eq:euler1}\\
\nabla \cdot v = 0.\label{eq:euler2}
\end{align}
The Euler equations depend on the fluid's velocity but not the spatial labels of fluid elements.  This ``particle-relabeling'' symmetry gives an infinite number of conserved quantities, by Noether's theorem \cite{arnold1966geometrie,arnold1969hamiltonian,arnold1999topological,marsden2002introduction,khesin2008geometry}.  The conserved quantities are the components of the fluid's convected momentum. They form an infinite-dimensional algebra with respect to the Lie-Poisson bracket:
\beq
\{J(\xi),J(\rho)\} = J([\xi,\rho]).\label{eq:algebra}
\eeq
There is a conserved charge, $J(\xi)$, for each divergence-free vector field, $\xi$.  Eq. \eqref{eq:algebra} corresponds to the superrotation part of the black hole horizon algebra \eqref{eq:q1}.  The main result of this section will be to derive \eqref{eq:algebra} using Noether's theorem.  

\subsection{The vortex}
\label{sec:vortex}

It will be helpful to have an example to refer back to.  A simple solution of the two-dimensional Euler equations is the vortex, with velocity and pressure
\begin{align}
v^i &= (-x_2,x_1),\label{eq:vortexv}\\
p &= \frac{1}{2}(x_1^2+x_2^2).
\end{align}
The motion of the fluid parcels is 
\begin{align}
x_1  &= \cos(t) X_1 - \sin(t) X_2\label{eq:x1}\\
x_2  &= \sin(t) X_1 + \cos(t) X_2.\label{eq:x2}
\end{align}
The $X_i$ are Lagrangian coordinates and the $x_i$ are Eulerian coordinates.  They correspond to the initial and current positions of the fluid parcels, respectively.   The map $x=\varphi(t,X)=\varphi_t(X)$ given by \eqref{eq:x1}-\eqref{eq:x2} defines the fluid's configuration. Differentiating with respect to time gives
\begin{align}
V^1 &= \dot{\varphi}_1 = -\sin(t) X_1 - \cos(t) X_2,\\
V^2 &= \dot{\varphi}_2 =\cos(t) X_1 - \sin(t) X_2.
\end{align}
$V(t,X)$ is the Lagrangian velocity and $v(t,x)$ is the Eulerian velocity.  We recover the Eulerian velocity \eqref{eq:vortexv} from the Lagrangian velocity using the change of coordinates $v=V\circ \varphi^{-1}$.

The Lagrangian and Eulerian velocities are measured with respect to a fixed spatial reference frame.  Another possibility is to measure velocity with respect to a comoving frame.  This defines the convected velocity
\beq
\calv \equiv \varphi^* V = \frac{\partial X^i}{\partial x^j} V^j= (-X_2,X_1)\label{eq:vortexcalv}.
\eeq
We pass to Lagrangian, Eulerian, and convected momenta by ``lowering indices'' on the velocities using the metric.  In this example the metric is flat, so the distinction between velocities and momenta is trivial.  

Now note that the convected momentum \eqref{eq:vortexcalv} is a constant independent of time.  The content of Noether's theorem for incompressible fluids is that the convected momentum is always a constant independent of time\footnote{possibly up to a total derivative, as we will soon see.}.  As a result, we get a conserved quantity,
\beq
J(\xi) = \int_{\mathbb{R}^2} \calv \cdot \xi \thinspace d^2X,
\eeq 
for each divergence-free\footnote{The reason for the divergence-free constraint will become clear later.  It is related to the fact that $\calv$ need only be conserved up to a total derivative.} vector field, $\xi$.  The underlying symmetry is the fact that the Euler equations only depend on the fluid's velocity and not on its configuration, $\varphi(X)$.  We will review the proof of this statement shortly.

In some sense, the vortex is too simple an example because the Eulerian momentum \eqref{eq:vortexv} also happens to be conserved.  We get a slightly more interesting example by passing to the boosted vortex, defined by
\begin{align}
v &\rightarrow v(t,x-ct) + c,\\
p &\rightarrow p(t,x-ct),
\end{align}
where $c$ is a constant vector.  The boosted vortex is also a solution of the Euler equations. The Eulerian momentum of the boosted vortex,
\beq
v=(-(x_2-c_2t)+c_1,(x_1-c_1 t)+c_2),
\eeq
is time dependent, but the convected momentum,
\beq
\calv = (-X_2,X_1),
\eeq
is conserved.

\subsection{General formulation of fluid dynamics}

Now consider the incompressible Euler equations on a $d$-dimensional manifold, $M$.  The configuration of the fluid is a volume-preserving diffeomorphism $\varphi\in Q\equiv\sdiff(M)$.  The volume-preserving condition encodes the incompressibility constraint.  The map $x=\varphi(X)$ takes Lagrangian coordinates to Eulerian coordinates.  A fluid flow is a time-dependent diffeomorphism, $\varphi(t,X)$.  The Lagrangian velocity is the time derivative
\beq
V=\dot{\varphi}(t,X) \in T_\varphi Q.
\eeq
We identify the space of $(\varphi,V)$ with the tangent bundle, $TQ$, and the fluid's phase space with the cotangent bundle, $T^*Q$.   For $\zeta\in TQ$ and $\alpha\in T^*Q$, we have the pairing
\beq
\langle\alpha,\zeta \rangle = \int_M \alpha\cdot \zeta\thinspace {\rm vol}_M.
\eeq

The Euler equations define a Hamiltonian flow on $T^*Q$.  The group $Q$ acts on itself by left and right translations, and these actions lift to actions on $T^*Q$.  The fluid's Hamiltonian is right invariant.  This corresponds to the particle-relabeling symmetry described above and it implies conservation of convected momentum.

\subsection{Right-invariance of fluid dynamics}

The actions of $Q$ on itself by left and right translations are
\begin{align}
L_\eta \varphi &= \eta\circ \varphi,\\
R_\eta \varphi &= \varphi\circ\eta,
\end{align}
where $\varphi,\eta\in Q$.  These actions lift to actions on the tangent bundle.   Let $\zeta\in T_\varphi Q$ and let $\gamma_\zeta(t)$ be an integral curve of $\zeta$ with $\gamma_\zeta(0)=\varphi$.  The lifted actions are
\begin{align}
TL_\eta \zeta &= \frac{d}{dt}\Big\vert_{t=0}L_\eta \gamma_\zeta (t)
	= \frac{d}{dt}\Big\vert_{t=0} \eta \circ \gamma_\zeta (t)
	= \eta_* \zeta,\\
TR_\eta \zeta &= \frac{d}{dt}\Big\vert_{t=0}R_\eta \gamma_\zeta (t)
	= \frac{d}{dt}\Big\vert_{t=0} \gamma_\zeta (t) \circ \eta
	= \zeta \circ \eta.
\end{align}
Note that $TL_\eta \zeta \in T_{\eta\circ\varphi}Q$ and $TR_\eta \zeta \in T_{\varphi\circ\eta}Q$.
These actions also lift to the cotangent bundle, $T^*Q$.  Let $\alpha\in T^*Q$ and $\zeta\in TQ$.  The cotangent lifts are defined by
\begin{align}
\langle T^*L_\eta \alpha,\zeta \rangle &= \langle \alpha,TL_\eta\zeta \rangle
	= \langle \alpha,\eta_*\zeta \rangle
	= \langle \eta^*\alpha,\zeta \rangle,\\
\langle T^*R_\eta \alpha,\zeta \rangle &= \langle \alpha,TR_\eta\zeta \rangle
	= \langle \alpha,\zeta\circ \eta \rangle
	= \langle \alpha\circ \eta^{-1},\zeta \rangle.
\end{align}
That is,
\begin{align}
T^*L_\eta \alpha &= \eta^* \alpha,\\
T^*R_\eta \alpha &= \alpha\circ\eta^{-1}.
\end{align}
Note that if $\alpha\in T^*_\varphi Q$, then $T^*L_\eta \alpha \in T^*_{\eta^{-1}\circ\varphi}Q$ and $T^*R_\eta \alpha \in T^*_{\varphi\circ\eta^{-1}}Q$.  
The cotangent lift of $R_\eta$ is a left action: $T^*R_{\eta\circ\lambda} = T^*R_\eta \circ T^*R_\lambda$.   The right lift of $R_\eta$ is defined by
\beq
R^*_\eta\alpha  = T^*_{\varphi\circ\eta} R_{\eta^{-1}} \alpha = \alpha\circ \eta.\label{eq:rlift}
\eeq
It is a right action: $R^*_{\eta\circ\lambda}=R^*_\lambda\circ R^*_\eta$.  It will prove more convenient to work with $R^*$ rather than $T^*R$.

Now consider the Hamiltonian $H:T^*Q\rightarrow \mathbb{R}$ defined by
\beq
H(\varphi,\alpha) = \frac{1}{2}\int_M v(\varphi,\alpha)^2 \thinspace{\rm vol}_M,
\eeq
where $ v(\varphi,\alpha) = \alpha \circ \varphi^{-1}$ is the Eulerian momentum.  The equations of motion are the Euler equations.  The right lift, $R^*_\eta$, sends $\alpha$ to $\alpha \circ \eta$ and $\varphi$ to $\varphi\circ \eta$.   So $ v= \alpha \circ \varphi^{-1}$ is invariant.   It follows that the Hamiltonian itself is right invariant:
\beq
H\circ R^*_\eta = H,\label{eq:rinv}
\eeq
for all $\eta\in Q$.  This is the particle-relabeling symmetry of the Euler equations described above.  It expresses the fact that fluid dynamics is independent of the fluid's configuration, $\varphi(X)$.

\subsection{Noether's theorem}

The conserved charges follow from Noether's theorem.  Let $\mathfrak{q}=\svect(M)$ be the Lie algebra of $Q$.  It is the algebra of divergence-free vector fields on $M$.  The infinitesimal version of \eqref{eq:rinv} is
\beq
\xi_P(H) = 0,
\eeq
where 
\beq
\xi_P = \frac{d}{dt}\Big\vert_{t=0}R^*_{\exp(t\xi)}
\eeq
is the infinitesimal generator of $R^*_\eta$ corresponding to $\xi\in \mathfrak{q}$.  $\xi_P$ is a vector field on $P=T^*Q$.

Our goal is to find a Hamiltonian vector field, $X_{J(\xi)}$, such that $\xi_P=X_{J(\xi)}$.  Conservation of $J(\xi)$ then follows from the Hamiltonian version of Noether's theorem:
\beq
\{H,J(\xi)\} = X_{J(\xi)}H = \xi_P(H) = 0.
\eeq
That is, we want $J(\xi)$ such that
\beq
\iota_{\xi_P}\Omega = dJ(\xi),
\eeq
where $\Omega=-d\Theta$ is the symplectic form on $T^*Q$.  Integrating gives
\beq
J(\xi) = \iota_{\xi_P}\Theta = \Theta(\xi_P),\label{eq:reduce}
\eeq
where $\Theta$ is the canonical one-form on $T^*Q$.  We have used Cartan's formula and the fact that $\xi_P$ preserves the canonical one-form\footnote{see proposition 6.3.2 of \cite{marsden2002introduction}.}: $\lie_{\xi_P} \Theta = 0$.

In finite dimensions, the canonical one-form is $\Theta = p_i dq^i$ and a vector field on phase space  has the coordinate expression $\chi=a^i \partial_{q^i}+b_i\partial_{p_i}$.  The action at $\alpha=p_i dq^i \in T^*Q$ is
\beq
\Theta_\alpha(\chi) = p_i a^i.
\eeq
To go to infinite dimensions, we need a coordinate-independent version of the right-hand side.  It is something like ``$\langle\alpha,\chi \rangle$,'' except the inner product as written cannot make sense, because $\alpha\in T^*Q$ while $\chi\in TT^*Q$.  However, the cotangent bundle comes equipped with a projection map $\pi_Q:T^*Q\rightarrow Q$, which lifts to a projection map $T\pi_Q:TT^*Q\rightarrow TQ$.  (In finite dimensions, $T \pi_Q$ acts simply by ``forgetting'' the $\partial_{p_i}$ piece of the vector field.)  This lets us build
\beq
\Theta_\alpha(\chi) = \langle \alpha,T\pi_Q(\chi) \rangle,
\eeq
which defines the canonical one-form in infinite dimensions.

Returning to the computation of the conserved charges, we have
\beq
J(\xi) = \Theta(\xi_P) = \langle\alpha,T\pi_Q(\xi_P) \rangle 
	= \langle \alpha, \xi_Q(\pi_Q\alpha)\rangle.
\eeq
In the last step, we used the fact that computing the infinitesimal generator on $P$ and then projecting down to a vector field on $Q$ gives the same result as projecting $P$ to $Q$ and then computing the infinitesimal generator.

The infinitesimal generator of right translations on $Q$ is
\beq
\xi_Q(\varphi) = \frac{d}{dt}\Big\vert_{t=0} \varphi\circ \exp(t\xi) = \varphi_* \xi.
\eeq
So we get our final expression for the conserved charges:
\beq
J(\xi) = \langle\alpha,\varphi_*\xi \rangle 
	= \langle \varphi^*\alpha,\xi \rangle 
	= \int_M \calv \cdot \xi\thinspace {\rm vol}_M.\label{eq:Jnoether}
\eeq 
$\calv=\varphi^*\alpha$ is the convected momentum corresponding to the Lagrangian momentum $\alpha$.  We get a conserved $J(\xi)$ for each $\xi\in \mathfrak{q}$.  It follows that the convected momentum itself is conserved, at least up to a total derivative.

\begin{figure}
\centering
\includegraphics[width=0.4\textwidth]{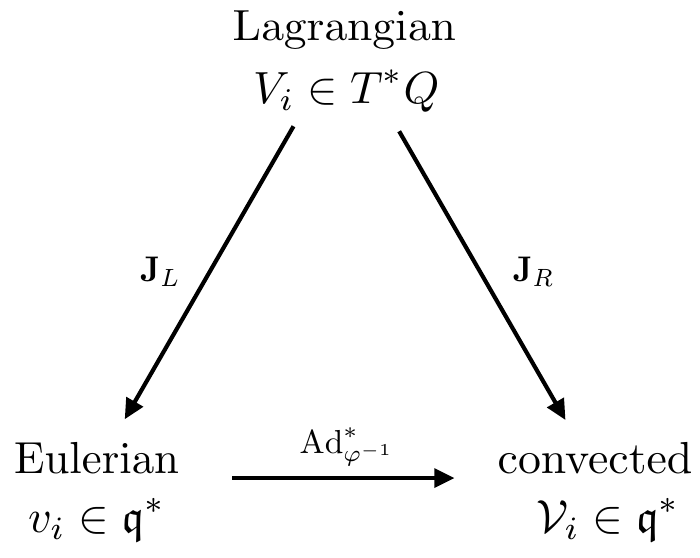}
\caption{The Lagrangian, Eulerian, and convected momenta are related by a triangle of maps.}
\label{fig:triangle}
\end{figure}

The map $\mathbf{J}_R:T^*Q\rightarrow \mathfrak{q}^*$ given by
\beq
\mathbf{J}_R(\alpha) = \varphi^*\alpha.
\eeq
is the right moment map.  In a similar way, by considering the infinitesimal generator of the cotangent lift of left translations, we arrive at the left moment map,
\beq
\mathbf{J}_L(\alpha) = \alpha\circ \varphi^{-1},
\eeq
which is the Eulerian momentum.   Right-invariant dynamics such as fluid dynamics conserves the right moment map, while left-invariant dynamics (such as rigid-body motion) conserves the left moment map.  The Eulerian and convected momenta are themselves related by the coadjoint action of $Q$ on $\mathfrak{q}^*$.  The relationships between the Lagrangian, Eulerian, and convected momenta can be summarized by the triangle in figure \ref{fig:triangle}.

\subsection{Charge algebra}

$\mathfrak{q}^*$ is a Poisson manifold with respect to the Lie-Poisson bracket.  Let $\calv\in \mathfrak{q}^*$ and let $F(\calv)$, $G(\calv)$ be functions on $\mathfrak{q}^*$.  The Lie-Poisson bracket is
\beq
\{F,G\}(\calv) = \left\langle \calv,\left[\frac{\delta F}{\delta \calv},\frac{\delta G}{\delta \calv}\right] \right\rangle,\label{eq:liepoisson}
\eeq
where $[,]$ is the vector field commutator\footnote{It is related to the Lie bracket on $\mathfrak{q}$ by a minus sign.}.  The Lie-Poisson bracket is the restriction of the canonical Poisson bracket, $\{,\}_P$, on $T^*Q$ to $\mathfrak{q}^*$:
\beq
\{F,G\}\circ \mathbf{J}_R  = \{F\circ \mathbf{J}_R,G\circ\mathbf{J}_R\}_P,
\eeq
Using eq. \eqref{eq:liepoisson}, we compute
\beq
\{J(\xi),J(\rho)\} = J([\xi,\rho]).\label{eq:algebra2}
\eeq
This is the algebra \eqref{eq:algebra} described in the introduction to this section.

\bibliography{ms}

\end{document}